\documentclass[epj]{svjour}
\usepackage{graphics}
\usepackage{amsmath}
\begin{document}
\title{Stress transmission in wet granular materials}
\author{V.~Richefeu\thanks{\emph{Present address:} richefeu@lmgc.univ-montp2.fr} 
\and F.~Radja\"i \and M. S.~El~Youssoufi
}
%
%
\institute{LMGC, UMR CNRS 5508, Cc. 048, Universit\'{e} Montpellier 2,\\ 
Place Eug\`{e}ne Bataillon, 34095 Montpellier Cedex 5, France}
\date{Received: date / Revised version: date}
%
\abstract{
We analyze stress transmission in wet granular media in the pendular
state by means of three-dimensional molecular dynamics simulations. We
show that the tensile action of capillary bonds induces a self-stressed
particle network organized in two percolating ``phases'' of positive and
negative particle pressures. Various statistical descriptors of the
microstructure and bond force network are used to characterize this
partition. Two basic properties emerge: 1) The highest  particle
pressure is located  in the bulk of each phase; 2) The lowest pressure
level occurs at the  interface between the two phases, involving also
the largest connectivity of the particles via tensile and compressive
bonds. When a confining pressure is applied, the number of tensile bonds
falls off and the negative phase breaks into aggregates and isolated
sites. 
} 
\maketitle

\section{Introduction}

The particle-scale origins of the strength and flow properties of dry
granular materials has been a subject of intensive research  since several
decades. It is now generally admitted that the scale-up of particle
interactions to the macroscopic scale is more subtle than initially expected
because of mediation by a disordered  microstructure  with a rich
statistical behavior \cite{Herrmann98,Roux01}. Considerable work has thus been devoted to
characterize the microstructure and its manifestations in the form of
highly inhomogeneous distributions of interparticle forces and
fluctuating particle velocities 
\cite{Oda72,Christoffersen81,Rothenburg89,Jaeger96,Calvetti97,Mueth98,Kuhn99,Roux00,Troadec02,Radjai02,Yang02}.
The bimodal character of the force
network \cite{Radjai98}, exponential probability functions of strong forces (force
chains) \cite{Coppersmith96,Radjai96,Mueth98,Majmudar05}, and clustering of dissipative contacts are recent examples of
this non trivial phenomenology \cite{Staron02}. Insightful analogies have also emerged
with other fields such as jamming transition in colloidal matter \cite{Cates98,Liu01,Dauchot05,Corwin05},
slow relaxation in glassy materials \cite{Knight95}, and fluid turbulence \cite{Radjai02}.

Most of our present knowledge on the subject excludes, however, cohesive
bonding between particles. Although one expects strong similarities due
to the common granular microstructure, the presence of 
cohesion brings about new mechanisms that tend to transform the nature
of the problem. At the macroscopic scale, the shear strength needs to be
described in terms of the Coulomb cohesion in addition to the internal
angle friction \cite{Wood90}. On the other hand, the interplay between cohesive bonds,
friction and rotation frustration of the particles leads to novel
features such as particle aggregation that control static and dynamic
properties of the material \cite{Castellanos01,Fillot06}. A well-known example is the wet sand where
small amounts of water affect significantly the bulk behavior \cite{Hornbaker97,Halsey98,Richefeu06}.
The phenomena arising from cohesion are of particular interest to the
processing (compaction, granulation\ldots) of fine powders \cite{Reynolds04,Castellanos05}. It seems
thus that a systematic investigation of the microstructure in cohesive
granular media should open new scopes for modeling granular materials. 

In this paper, we analyze the force network in wet granular assemblies
of spherical particles. We are interested in a basic question: how
cohesive grains keep together to form a self-sustained structure in the
absence of confinement (no container and no confining stresses)? The
packing  can reach an equilibrium state as a result of attraction forces
and elastic repulsion between particles without or with self-stressed
structures. While the particles are balanced in both cases, the
attraction force at each contact is exactly balanced by an elastic
repulsion force in the first case. In contrast, in the second case  all
contacts are not in their equilibrium state due to steric hinderance
between particles. Hence, a network of tensile and compressive forces is
formed inside the packing. These self-equilibrated forces can be induced
through various loading histories such as consolidation \cite{Preecha02,Radjai01} 
or differential particle swelling \cite{ElYous05}. In
wet granular media in the pendular state, the self-stresses appear
naturally due to the tensile action of capillary bonds bridging the gaps
between neighboring particles within a debonding distance. We focus in
this paper on the structure of these self-stresses induced by capillary
bonds.

We use 3D molecular dynamics method in which capillary attraction
between spherical particles is implemented as a force law expressing the
capillary force as a function of the distance, water volume, and
particle diameters. The total water volume is distributed homogeneously
between particles. The packing is analyzed at equilibrium under zero
confining pressure. The main goal of our analysis is to characterize the
organization of particle pressures which take positive or negative
values according to their positions in the network of self-equilibrated
bond forces. We will see that this organization involves a genuin
partition of the particles in two phases of negative and positive
pressures. In the following, we first describe the simulation method and
our model of capillary cohesion. In Sections~\ref{sec:3} and
\ref{sec:4}, we study in detail the probability density functions of the
forces and the connectivity of particles via tensile and compressive
bonds. Then, in Sections \ref{sec:5} and \ref{sec:6} we introduce
particle stresses and we analyze their distribution and correlation with
the connectivity of the particles. Section \ref{sec:7} is devoted to the
influence of external pressure. We conclude with a discussion about the
main findings of this work.

\section{Numerical method \label{sec:2}}

\begin{figure}
\centering
\resizebox{0.95\columnwidth}{!}{%
  \includegraphics{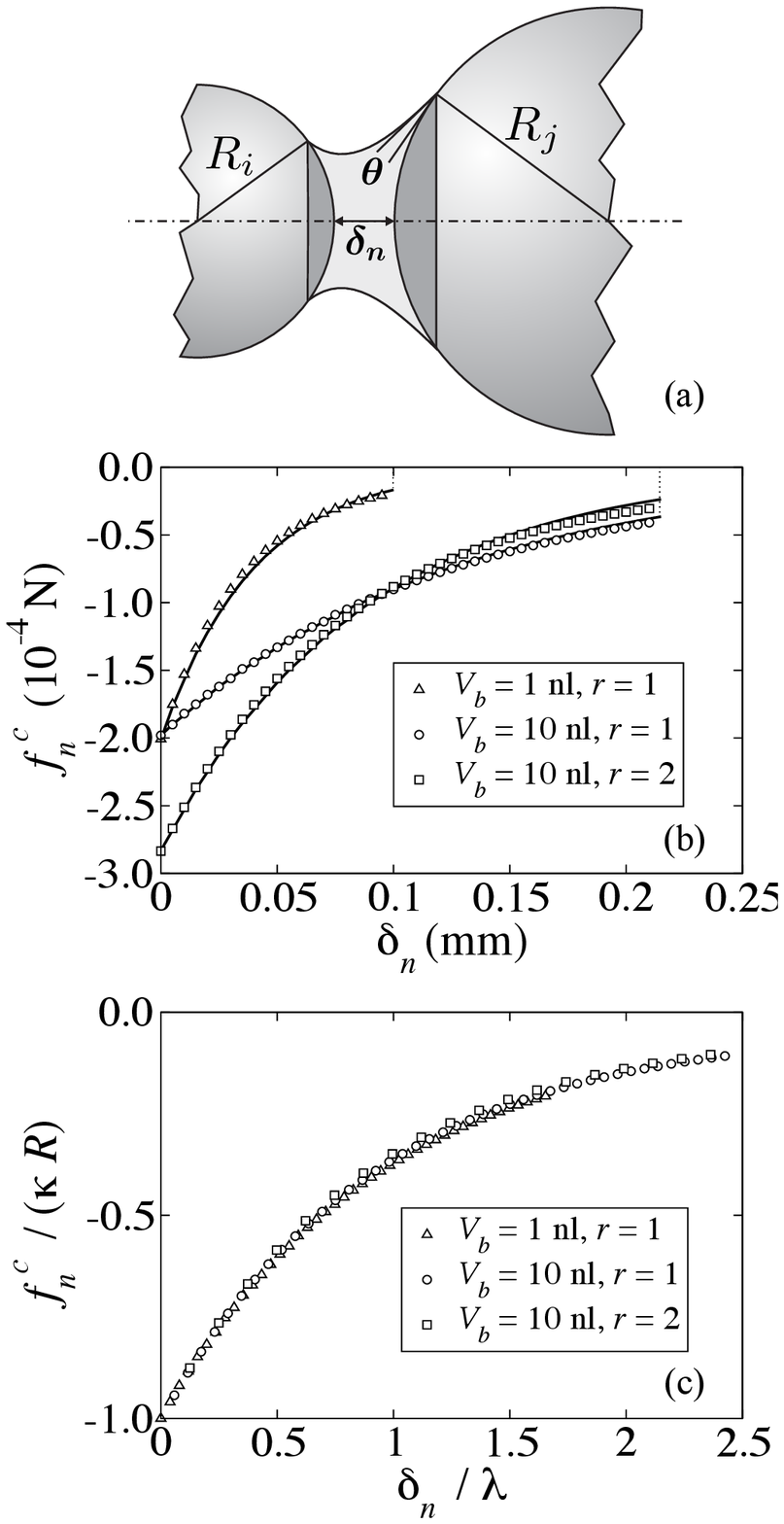}
}
\caption{
(a) Geometry of a capillary bridge; 
(b) Capillary force $f_n^c$ as a function of the gap $\delta_n$
between two particles for different
values of the liquid volume $V_b$ and size ratio $r$ according to the 
model proposed in this paper (solid lines), and from
direct integration of the Laplace-Young equation (open circles); 
(c) Scaled plot of the capillary force as a function of gap from the direct data shown in (b). 
}
\label{fig:1}
\end{figure}

We used the molecular dynamics (MD) method with a velocity Verlet integration
scheme \cite{Cundall79,Allen87}. The force laws involve normal
repulsion, capillary cohesion, Coulomb friction, and normal damping. The
normal force has three different sources,  
\begin{equation} \label{eq:decomp}
f_n = f_n^e + f_n^d + f_n^c .
\end{equation} 
The first term is the repulsive contact force depending linearly on the 
normal distance $\delta_n$ between the particles (Fig~\ref{fig:1}(a)):
\begin{equation} \label{eq:ctc-law}
f_n^e = 
\left \{
\begin{array}{ll}
-k_n \delta_n & \quad \mbox{for} \quad \delta_n < 0 \\
0             & \quad \mbox{for} \quad \delta_n \geq 0 
\end{array}
\right.,
\end{equation}
where $k_n$ is the normal stiffness. The second term represents a
viscous damping force depending on the normal velocity $\dot{\delta}_n$:
\begin{equation} \label{eq:damping-law}
f_n^d = 
\left \{
\begin{array}{ll}
2 \alpha_n \sqrt{m k_n}\; \dot{\delta}_n & \quad \mbox{for} \quad \delta_n < 0 \\
0             & \quad \mbox{for} \quad \delta_n \geq 0
\end{array}
\right.,
\end{equation}  
where $m=m_im_j/(m_i+m_j)$ is the reduced mass of the particles $i$ and
$j$, $\alpha_n$ is a damping rate varying in the range $[0,\;1[$ and
that accounts for the rate of normal dissipation or the restitution
coefficient between particles \cite{Shafer96}.

The last term in Eq.~\ref{eq:decomp} is the capillary force depending on
the liquid bond parameters, namely the gap $\delta_n$, the liquid bond
volume $V_b$, the liquid surface tension $\gamma_s$, and the
particle-liquid-gas contact angle $\theta$. The capillary force can be
obtained by integrating the Laplace-Young equation \cite{Erle71,Lian93,Mikami98,Soulie06}. 
However, for
molecular dynamics simulations, we need an explicit expression
of $f_n^c$ as a function of the liquid bond parameters. 
We propose the following simple form:
\begin{equation} \label{eq:coh-law}
f_n^c = 
\left \{
\begin{array}{ll}
- \kappa\ R & \quad \mbox{for} \quad \delta_n < 0 \\
- \kappa\ R\ e^{-\delta_n / \lambda} & \quad \mbox{for} \quad 0 \leq \delta_n \leq \delta_n^{max} \\
0 & \quad \mbox{for} \quad \delta_n > \delta_n^{max} \\
\end{array}
\right. .
\end{equation}
where $R=\sqrt{R_i R_j}$ is the geometrical mean of particle radii,
and $\lambda$ is a length scale to be discussed below. 
The prefactor $\kappa$ is given by \cite{Willett00,Bocquet02,Herminghaus05}
\begin{equation}
\kappa =  2 \pi \gamma_s \cos \theta ,
\end{equation}
and $\delta_n^{max}$ is the debonding distance given by \cite{Lian98}
\begin{equation} \label{eq:debond-dist}
\delta_n^{max} = \left ( 1+ \frac{\theta}{2} \right ) V_b^{1/3}.
\end{equation}
The capillary bridge is stable as long as $\delta_n < \delta_n^{max}$.
In the simulations, the bridge is removed
as soon as the debonding distance is reached, and the liquid is
redistributed among the contacts belonging to the same particle 
in proportion to grain sizes \cite{Richefeu06}.
At contact ($\delta_n = 0$), $\kappa R$ corresponds 
to the largest attraction force between
two particles. In the simulations presented in this paper, 
we assume that particles are perfectly 
wettable, i.e. $\theta = 0$. This is a good approximation for 
water and glass beads.

The length $\lambda$ governs the exponential falloff of the capillary force
in Eq.~\ref{eq:coh-law}. It should depend on the liquid volume $V_b$, a mean radius $R'$, and the ratio
$r\!=\!\max \{ R_i/R_j ; R_j/R_i\}$. A reasonable choice is 
\begin{equation}  \label{eq:lambda_choice}
\lambda = c\ h(r) \left ( \frac{V_b}{R'} \right )^{1/2},
\end{equation}
where $c$ is a constant prefactor, $h$ is a function depending on the ratio $r$,
and $R'$ is the harmonic mean ($R'= 2R_iR_j/(R_i+R_j)$).
When introduced in Eqs.~\ref{eq:lambda_choice} 
and \ref{eq:coh-law}, this form yields a nice fit for the capillary force
obtained from direct integration of the Laplace-Young equation by simply 
setting $h(r) = r^{-1/2}$ and $c \simeq 0.9$.
Fig.~\ref{fig:1}(b) shows the plots of Eq.~\ref{eq:coh-law} for different
values of the liquid volume $V_b$ and size ratio $r$
together with the corresponding data from direct integration. 
The fit is excellent for 
$\delta_n = 0$ (at contact) and for small gaps.

Figure \ref{fig:1}(c) shows the same plots of the direct data as in Fig.~\ref{fig:1}(b)
but in dimensionless units where the forces are normalized
by $\kappa R$ and the lengths by $\lambda$. 
We see that the data collapse indeed on the same plot, indicating again that
the force $\kappa R$ and the expression of $\lambda$ in Eq.~\ref{eq:lambda_choice}
characterize correctly the behavior of the capillary bridge for $\theta = 0$.
From the same data, we also checked
that the geometric mean  $R=\sqrt{R_i R_j}$ introduced in Eq.~\ref{eq:coh-law}
provides a better fit than the harmonic mean $2R_iR_j/(R_i+R_j)$ proposed 
by Derjaguin for polydisperse particles in the limit of small gaps \cite{Israelachvili92}.

For the friction force $\vec{f}_t$, we used a viscous-regularized
Coulomb law \cite{Shafer96,Dippel97,Luding98}
\begin{equation} \label{eq:coulomb-law}
\vec{f}_t=-\min \left \{
\gamma_t ||\dot{\boldsymbol{\delta}}_t||,\; \mu (f_n - f_n^c)
\right \} 
\frac{\dot{\boldsymbol{\delta}}_t}{||\dot{\boldsymbol{\delta}}_t||} ,
\end{equation} 
where $\gamma_t$ is a tangential viscosity parameter, $\mu$ is the
coefficient of friction, and $\dot{\boldsymbol{\delta}}_t$ is the
sliding velocity. 
In relaxation to  equilibrium, $\dot{\boldsymbol{\delta}}_t$ declines 
but never vanishes due to residual kinetic energy. The equilibrium 
state is practically reached as soon as we have 
$\gamma_t ||\dot{\boldsymbol{\delta}}_t|| < \mu (f_n - f_n^c)$
at all contacts. This means that the friction force is inside the
Coulomb cone everywhere in the system.

\begin{table}[h]
\caption{Parameters  used in the simulations.}
\begin{center}
\begin{tabular}{llll}
\hline
Parameter & Symbol & Value & Unit \\
\hline
Normal stiffness          & $k_n$      & $1000$    & N/m  \\
Damping rate              & $\alpha_n$ & $0.8$     & -    \\
Tangential viscosity      & $\gamma_t$ & $1$       & Ns/m \\
Capillary force prefactor & $\kappa$   & $0.4$     & N/m  \\
Gravimetric water content &            & $0.007$   & -    \\
Liquid density            &            & $1000$    & kg.m$^{-3}$ \\
Particle density          &            & $2700$    & kg.m$^{-3}$ \\
Friction coefficient      & $\mu$      & $0.4$     & -    \\
Time step                 &            & $10^{-6}$ & s    \\ 
\hline 
\end{tabular}
\end{center}
\label{tab:1}
\end{table}

The simulations were performed with a packing composed of $N = 8000$
spheres with diameters $d=1$, $1.5$, and $2$ mm, in equal numbers. The
system was subjected to an isotropic pressure $p_m$ via six rigid walls
and no gravity in order to obtain an as homogeneous state as possible.
For the same reason, the friction with the walls was set to zero although
wall effects can not be fully removed. The parameters used in the
simulations are displayed in Table~\ref{tab:1}. The choice of the water 
volume has no influence on the value of the largest capillary force in the 
pendular state. We also note that the liquid bonds are homogeneously 
distributed into all gaps within the debonding distance \cite{Richefeu05b}.
With a gravimetric water content of $0.007$, the coordination number 
is $\simeq 6$. 
Experiments show that the distribution of liquid bonds
depends on the preparation protocol involving the water volume, 
mixing procedure, and waiting times \cite{Fournier05,Richefeu06}.

\section{Force distributions \label{sec:3}}

\begin{figure}
\centering
\resizebox{0.95\columnwidth}{!}{%
  \includegraphics{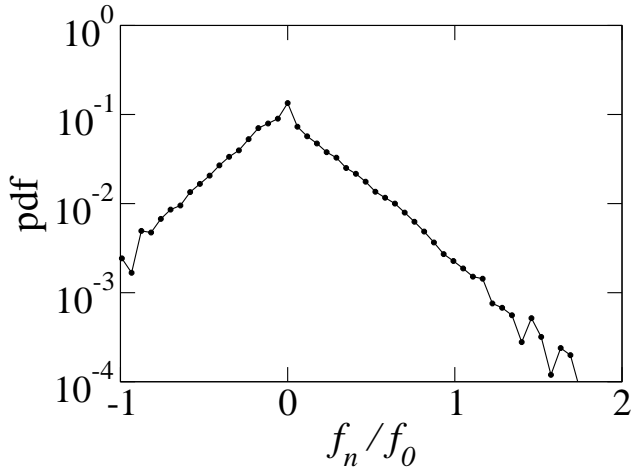}
}
\caption{Probability density function of normal forces
normalized by the largest capillary force $f_0$ for zero confining 
pressure.}
\label{fig:2}
\end{figure}

We start out by considering  force distributions first in a system
subjected to a negligibly small average compressive stress $p_m \simeq
0$~Pa. Fig.~\ref{fig:2} displays the probability density function (pdf) of 
the normal forces both
in  tensile (negative) and compressive (positive) ranges. We observe  a
distinct peak centered on $f_n = 0$ and two nearly symmetrical parts 
decaying exponentially from the center: 
\begin{equation}
P \propto e^{-\alpha |f_n|/f_0},
\end{equation}
with $\alpha \simeq 4$ within statistical precision for both negative
and positive forces, and $f_0 = \kappa R_{max}$, where $R_{max}$ is 
the largest particle radius. It has been observed that in dry granular media the
distribution is not purely exponential in the whole range of bond
forces \cite{Radjai96,Mueth98,Majmudar05,vanHecke05}. 
Below the average force, the distribution tends to be
uniform or a decreasing power law with a weak exponent \cite{Radjai99}. 
In the present case
of wet cohesive materials, the exponential behavior extends to the center 
of the distribution. 
It is important to remark that this peak does not
represent non-transmitting contacts. It rather corresponds to
contacts where capillary attraction is balanced by elastic
repulsion, i.e. $k_n \delta_n + f_0 = 0$. We will see below that this
peak persists under the action of a compressive confining stress, suggesting that
its presence reflects a feature of force transmission in wet granular
materials.

The tensile range is cut off  at $f_n = -f_0$ corresponding to the
largest capillary force. Although the confining stress is zero,
positive forces as large as $2 f_0$ can be found in the system.  
In contrast to dry
granular materials, the pdf shows a peak at $f_n = 0$ which is the
average force in the present case. In fact, in an unconfined assembly of
dry rigid particles, no self-stresses appear and the forces vanish
at all contacts. In our wet material, the presence of liquid bonds
induces tensile and compressive self-stresses although the average force
is zero.

\begin{figure}
\centering
\resizebox{0.85\columnwidth}{!}{%
  \includegraphics{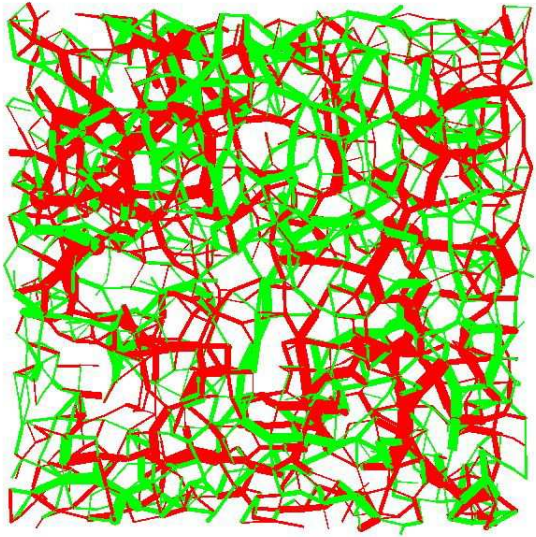}
}
\caption{(Color online) A map of tensile (green) and compressive (red)
 forces in a thin layer cut out in the packing. Line thickness is 
 proportional to the magnitude of the force.}
\label{fig:3}
\end{figure}

Figure \ref{fig:3} shows the force network in a narrow slice nearly
three particle diameters thick. The tensile and compressive 
forces are represented by segments of different colors 
joining particle centers. The line thickness is proportional to the force. It
is remarkable that tensile and compressive force chains can be observed
although the slice is quite narrow.  The bond coordination
number $z$ (average number of bonds per particle) is $\simeq 6.1$ including 
nearly $2.97$ compressive bonds and $3.13$ tensile bonds.

\section{Connectivity of the bond network \label{sec:4}}

\begin{figure}
\centering
\resizebox{0.7\columnwidth}{!}{%
  \includegraphics{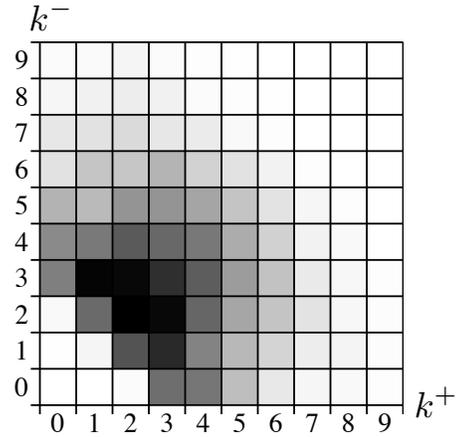}
}
\caption{Grey level map of the connectivity function $\rho(k^+,\,k^-)$.}
\label{fig:4}
\end{figure}

We analyze the connectivity of the particles via capillary bonds 
by considering the fraction $\rho (k^+,\,k^-)$ of particles
with exactly $k^+$ compressive bonds and $k^-$ tensile bonds. This
function is displayed in Fig.~\ref{fig:4} as grey level map in the
parameter space $(k^+,\ k^-)$ for our system. The map is
symmetric with respect to the line $k^+ = k^-$, reflecting thus the
symmetrical roles of compressive and tensile   networks in the absence of
confining stresses. A peak value of $\rho$ occurs at  $k^+ = k^- = 2$.
Obviously, the condition of particle equilibrium cannot be fulfilled
with $k^+ \leq 1$ and $k^- \leq 1$ and the corresponding levels are zero
on the map. The particles with  $k^+ = 2$ and $k^- = 0$
define chains of compressive bonds whereas  particles with  $k^+ = 0$
and $k^- = 2$ correspond to chains of tensile bonds. But such ``pure''
chains are rare. At larger values of $k^+$ and $k^-$ the fractions
decline basically due to geometrical hinderance between particles.

\begin{figure}
\centering
\resizebox{0.95\columnwidth}{!}{%
  \includegraphics{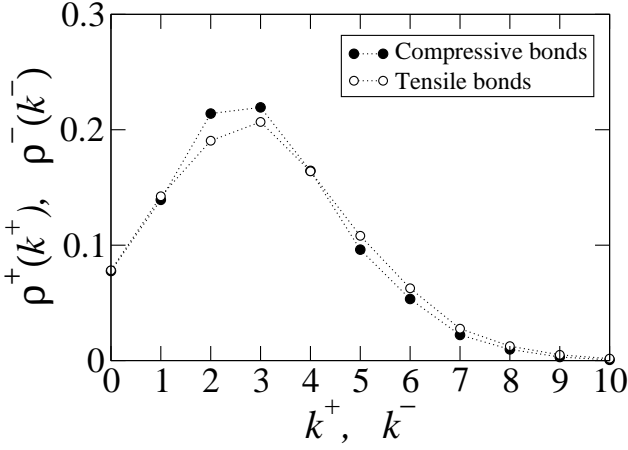}
}
\caption{Reduced connectivity functions at zero confining pressure for 
tensile and compressive bonds. 
}
\label{fig:5}
\end{figure}

An interesting feature of the connectivity map (Fig.~\ref{fig:4}) is the
existence of a population of particles involving no tensile bonds (the
row $k^- = 0$) as well as a population of particles involving no
compressive bonds (the column  $k^+ = 0$). Reduced connectivity
functions $\rho^+$ and $\rho^-$  can be defined by summing up  the
function $\rho (k^+,\,k^-)$ along columns and rows, respectively:
\begin{eqnarray}
\rho^+(k^+) = \sum_{k^-} \rho (k^+,\,k^-), \nonumber \\
\rho^-(k^-) = \sum_{k^+} \rho (k^+,\,k^-).
\end{eqnarray}
The plots of these functions are shown in Fig.~\ref{fig:5}. They nearly
coincide as expected from the symmetry observed in Fig.~\ref{fig:4}. We
have $\rho^+(0) = \rho^-(0)\simeq 0.08$, corresponding respectively to particles in
purely extensional or compressional local environments. A maximum occurs
at $k^+ = k^- = 2$ or $3$.

\section{Particle stresses \label{sec:5}}

Until  now, we focussed on  forces and their distributions with regard
to the tensile or compressive nature of the bonds. For the description
of stress transmission in our system we need, however, to characterize
the inhomogeneities at the scale of particles representing the smallest entities 
for which the equations of motion are resolved in MD simulations. At
this scale, a Cauchy stress in the sense of continuous media cannot be
defined. But, it can be shown that the so-called {\em internal moment
tensor} $\boldsymbol{M}$ has the same properties as the Cauchy
stress tensor $\boldsymbol{\sigma}$, and its definition extends
mathematically to discrete media at all scales down to the particle
scale \cite{Moreau97,Staron05}. For a particle $i$ subjected to forces
$\mathbf{f}^{ij}$ from its contact neighbors $j$ at the contact points
$\mathbf{r}^{ij}$, the internal moment tensor $\boldsymbol{M}_i$ is
given by \cite{Moreau97}
\begin{equation}
\boldsymbol{M}_i = \sum_{j \ne i} \mathbf{f}^{ij} \otimes \mathbf{r}^{ij}, 
\end{equation} 
where $\otimes$ designs a tensorial product. The internal moment tensor
is additive and independent of the origin of the coordinate frame. 
For a collection of particles in a
control volume $V$, the total internal moment $\boldsymbol{M}$ is simply
the sum of the particle moments:
\begin{equation}
\boldsymbol{M} = \sum_{i \in V} \boldsymbol{M}_i.
\end{equation}
This tensor has the dimension of a moment and it becomes homogeneous to
a stress when divided by the control volume:
\begin{equation}
\boldsymbol{\sigma} = \frac{1}{V} \sum_{i \in V} \boldsymbol{M}_i.
\label{eq:M}
\end{equation}
At large scales (containing a sufficiently large number of particles), 
the volume is well-defined and the stress tensor is equivalent to the
internal moment tensor divided by this volume. However, at the particle
scale, while $\boldsymbol{M}_i$ is  defined in a unique way, the choice
of the volume $V$ is a matter of convenience. It is
reasonable to take into account the {\em free volume} $V_i$ of each
particle $i$, as the sum of the volume of the particle and part of 
the surrounding pore
space. On average, we have $V_i = (1/6)\pi d_i^3 / \nu$,
where $d_i$ is the particle diameter and $\nu$ denotes the packing
fraction. With this choice, the particle stress tensor
$\boldsymbol{\sigma}_i$ takes  the following form: 
\begin{equation}
\boldsymbol{\sigma}_i = 6 \frac{\nu}{\pi d_i^3} \sum_{j \ne i} 
\mathbf{f}^{ij} \otimes \mathbf{r}^{ij}.
\label{eq:ParticleStress}
\end{equation}
Remark that when summing this form over many particles in a control
volume as in Eq.~\ref{eq:M}, each contact $ij$ enters the summation two
times, belonging once to particle $i$ and once to particle $j$. Since
$\mathbf{f}^{ij} = - \mathbf{f}^{ji}$, the contribution of the contact
$ij$ to stress is $\mathbf{f}^{ij} \otimes (\mathbf{r}^{ij} -
\mathbf{r}^{ji})$, a writing that involves the center-to-center vector
instead of position vectors of the contact points if the origin of coordinates 
for each particle is chosen coincides with the center of the particle. 
However, we consider
here only the particle stress, and at this scale, according to
Eq.~\ref{eq:ParticleStress}, only the position vectors of contact points
are involved.   

\begin{figure}
\centering
\resizebox{0.95\columnwidth}{!}{%
  \includegraphics{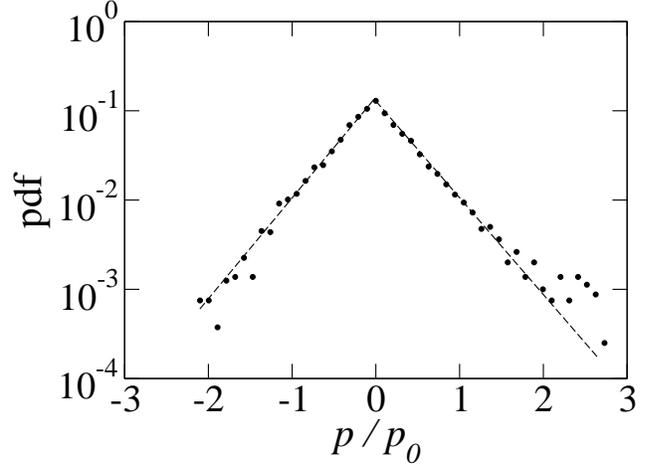}
}
\caption{Probability density function of particle pressures
normalized by reference pressure $p_0$ (see text) in the unconfined packing.}
\label{fig:6}
\end{figure}

Here, we explore the particle pressures (average particle stresses)
$p_i= (1/3)\,\mbox{tr}\,\boldsymbol{\sigma}_i$. Each  particle can take
on positive or negative pressures according to the nature of the forces
exerted by contact neighbors. The pdf of particle pressures is displayed
in Fig.~\ref{fig:6}. The pressures have been normalized by a reference
pressure $p_0 =  f_0 / \langle d \rangle^2 $. The
distribution is symmetric around and peaked on zero pressure, and each
part is well fit by an exponential form. Obviously, the exponential
shape of particle pressure distributions reflects statistically that of
bond forces. In dry granular media, since the normal forces are all of
the same sign (compressive) and particle pressure results from the
summation of individual bond forces on a particle, the probability is
expected to   vanish as the  pressure goes to zero. In
contrast, Fig.~\ref{fig:6} shows that the exponential shape of particle
pressure distribution  extends to the center of the pdf. This can be  related
to the fact that all normal forces are not of the same sign. 
On the other hand, the zero
pressure corresponds to a state where a particle is balanced under the
combined action of tensile and compressive forces. Since such particle states are
not marginal here, they might reflect a particular 
organization of the stresses in the wet packing (see below). 
    
\section{Partition of particle pressures \label{sec:6}} 

An interesting observation about the connectivity map (Fig.~\ref{fig:4})
was the presence of particles with only tensile or only compressive
bonds. In terms of particle pressures, these populations carry
negative or positive pressures, respectively. However, this information
is not rich enough as it does not specify the pressure levels in these
populations. The question is how the particle pressure
is locally correlated with the particle connectivity.  For particle $i$,
the connectivity is specified by the number $k^+_i$ of compressive bonds
and the number $k^-_i$ of tensile bonds. Let us now consider the set
${\cal S}(k^+,\,k^-)$ of particles  $i$ such that $k^+_i = k^+$ and
$k^-_i = k^-$. The {\em partial} pressure carried by this set is the sum of
particle pressures in this set divided by the total number of particles:
\begin{equation}
p(k^+,\,k^-) = \frac{1}{N} \sum_{i \in {\cal S}(k^+,\,k^-)} p_i .
\end{equation}
This is obviously an additive quantity so that the average stress $p_m =
\sum_{(k^+,\,k^-)} p(k^+,\,k^-)$. The function $p(k^+,\,k^-)$ provides
detailed information about the way particle pressures are
distributed with respect to the bond network. In other words, this
function describes the relationship between the pressure sustained by a
particle and its first neighbors with which it is bonded. 

\begin{figure}
\centering
\resizebox{0.7\columnwidth}{!}{%
  \includegraphics{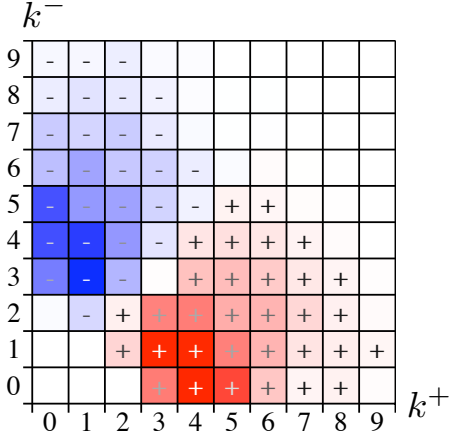}
}
\caption{(Color online) The map of partial pressures $p(k^+,\,k^-)$.}
\label{fig:7}
\end{figure}

Figure \ref{fig:7} shows the map of partial pressures  $p(k^+,\,k^-)$
in the parameter space $(k^+,\,k^-)$.
Interestingly, we observe a bipolar structure of partial pressures which
is antisymmetric with respect to the line $k^+ = k^-$ within 
statistical precision. The pressures are positive
in the range $k^+ > k^- $ and negative in the range $k^+ < k^- $. Hence,
the line $k^+ = k^-$ defines the transition zone between the two parts
with $p(k^+,\,k^-=k^+) \simeq 0$. This line corresponds to the largest
population of particles  according to the connectivity map
(Fig.~\ref{fig:4}). The pressure extrema are located at $(k^+ = 4,\,k^- =
0)$ for positive pressures and at $(k^+ = 0,\,k^- = 4)$ for negative
pressures.  

\begin{figure}
\centering
\resizebox{0.75\columnwidth}{!}{%
  \includegraphics{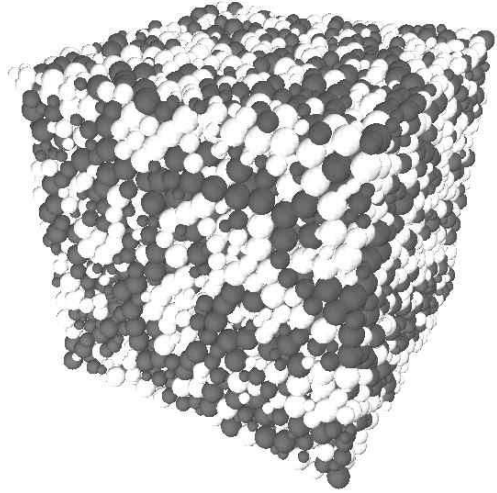}
}
\caption{A representation of the packing with particles of 
negative (white) and positive (black) pressures.}
\label{fig:8}
\end{figure}

The bipolar structure of the pressure map suggests that the particles of
positive and negative pressures define two separate  phases throughout
the system. In this picture, the population of particles along the line
$k^+ = k^-$ corresponds to the particles at the interface between the
two phases. This interpretation is backed by Fig.~\ref{fig:8} displaying
the packing where the two phases are represented in black and white.
The particles of either positive or negative pressure percolate 
throughout the system. The two
phases are intimately intermingled with a large interface between
them. The particles at the interface belong to the  line $k^+ = k^-$
corresponding to a fraction $0.125$ of particles. 
The morphology of the two phases is approximately
filamentary with variable thickness. 

\begin{figure}
\centering
\resizebox{0.95\columnwidth}{!}{%
  \includegraphics{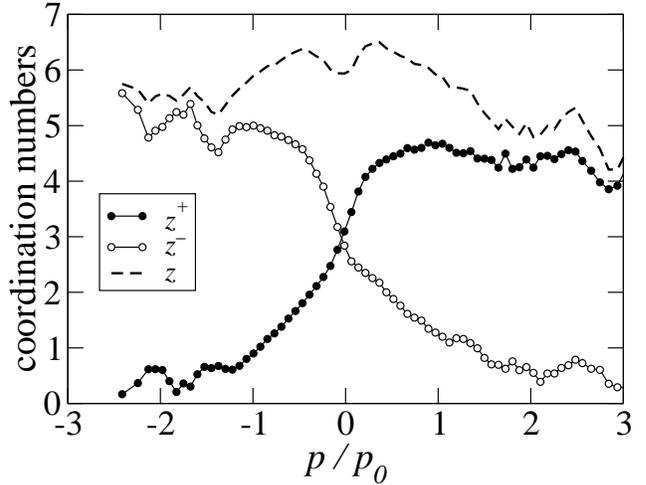}
}
\caption{Average numbers of
tensile ($z^-$) and compressive ($z^+$) bonds per particle as well as 
the partial coordination number ($z = z^- + z^+$) as
a function of the particle pressure in the unconfined packing.}
\label{fig:9}
\end{figure}

The correlation between the bond connectivity and particle pressures can
alternatively be determined by considering the average numbers of
tensile and compressive bonds per particle, denoted $z^-$ and $z^+$, as
a function of the particle pressure $p$. In order  to evaluate these
functions, we consider the set ${\cal S} (p)$ of particles $i$ with a
pressure $p_i$  in the range $[p,\,p+\Delta p]$, where $\Delta p$ is the
pressure increment. The partial coordination numbers  $z^-$ and $z^+$
are defined by 
\begin{eqnarray} 
z^+ (p)= \frac{1}{N(p)} \sum_{i \in {\cal S}(p)} k_i^+, \nonumber  \\ 
z^-  (p)= \frac{1}{N(p)} \sum_{i \in {\cal S}(p)} k_i^-, 
\end{eqnarray} 
where $N(p)$ is the number of
particles in the set. These functions are plotted in Fig.~\ref{fig:9} in
the case $p_m = 0$. The plot of   $z^- (p)$ is an approximate mirror
image of $z^+ (p)$ with respect to $p=0$. Three zones can be discerned.
For $p < -p_0$, we have $z^- \simeq 5$ and $z^+ \simeq 0.5$. This zone
represents mainly the particles belonging  to the {\em bulk} of the
negative phase.  For $p >p_0$, we have $z^- \simeq 0.5$ and $z^+ \simeq
4$. The particles in this zone belong  to the {\em bulk} of the positive
phase.  In the range $-p_0 < p < p_0$, $z^+$ increases and  $z^-$
declines. The intersection occurs at $p=0$ where $z^- = z^+ \simeq 3$.
This zone corresponds to the particles located at the interface between 
negative and positive phases.  

The above findings underline that stress  transmission in wet granular 
media is non trivial. 
In particular, they support the partition of the packing into two
well-defined phases both in terms of  particle connectivities and in
terms of particle pressures. The peculiarity of this partition is that
the extrema of particle pressures are located in the bulk of each phase
(Fig.~\ref{fig:4}) whereas the maximum of connectivity between particles
via tensile and compressive bonds resides at their interface
(Fig.~\ref{fig:7}). Along this interface, not only the bond forces are
balanced on each particle, as everywhere in the packing, but the tensile
and compressive bonds contribute equally in such a manner that the
particle pressures are extremely low.

\section{Influence of confining pressure \label{sec:7}}

In the absence of a confining stress, the capillary bonds are at the
origin of self-stresses or self-equilibrated forces that we analyzed in
preceding sections. Hence, the observed symmetry between tensile and
compressive bonds is a consequence of static equilibrium at zero
confining stress. The question remains whether the partition of particle
pressures, as depicted above, still holds when a wet packing is
subjected to (compressive) confining stress. In practice, however, we
cannot isolate self-equilibrated forces for each particle from those 
induced by the
external stress (the actual force network being the sum of the two
networks).
 This is because the external pressure drives the packing to a
new equilibrium state with modified microstructure. Hence, the
self-stresses for $p_m = 0$, i.e. before rearrangements, do not
correspond to the rearranged state for $p_m \neq 0$.
 While it can be conjectured that the
self-stresses in the presence of a confining pressure will display the
same ``bipolar'' behavior as for  $p_m = 0$, we consider here the force
network and particle pressures without distinction between induced and
self-equilibrated forces. 

We applied an isotropic
stress $p_m = 100$~Pa to the wet packing prepared with $p_m = 0$.   
The packing  was then allowed to
relax to equilibrium under the action of the applied pressure. 
This level of confinement is high compared to the
reference pressure $p_0$ ($p_m/p_0 \simeq 0.5$), yet not too high to
mask fully the manifestations of capillary cohesion. The same packing
was also compressed isotropically for  $p_m = 100$~Pa without capillary
cohesion (dry packing).

\begin{figure}
\resizebox{0.95\columnwidth}{!}{%
  \includegraphics{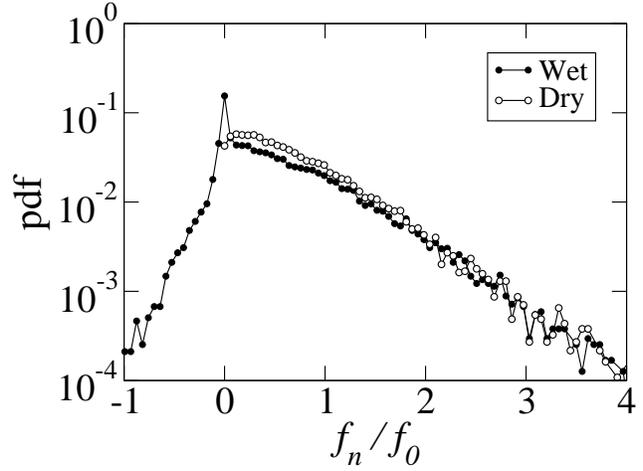}
}
\caption{Probability density function of normal forces
normalized by the largest capillary force $f_0$ in the confined packing.}
\label{fig:10}
\end{figure}

The pdf of normal forces is shown in Fig.~\ref{fig:10} for both 
packings. The symmetry of the distribution around $f_n = 0$ is now
broken (compare to Fig.~\ref{fig:2}). The distributions  are roughly
exponential for both tensile and compressive forces but the exponents
are different. We also note that the exponent for compressive forces is
nearly the same as for the dry sample. On the other hand, although the
strictly zero forces have been removed from the statistics, a distinct
peak centered on zero force is present for the wet sample and absent
from the dry sample. If the occurrence of this peak in the unconfined case
($p_m = 0$) is attributed to the interfacial zone, 
its persistence in the confined case
suggests that a negative pressure phase continues to coexist with the
positive pressure phase (which is now the dominant phase assisted with
the confining stress). This point will be analyzed below in relation  to
the distribution of particle pressures.

\begin{figure}
\centering
\resizebox{0.95\columnwidth}{!}{%
  \includegraphics{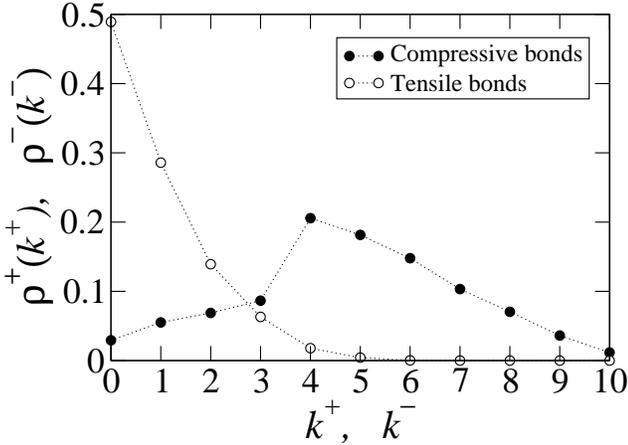}
}
\caption{Reduced connectivity functions in the confined packing for 
tensile and compressive bonds.}
\label{fig:11}
\end{figure}

The coordination numbers for compressive and tensile bonds are $4.85$
and $0.85$, respectively. This shows that the application of a confining
pressure has transformed a fraction of tensile bonds into compressive
bonds. The plots of the connectivity functions  $\rho^+ (k^+)$ and
$\rho^- (k^-)$ are displayed in Fig.~\ref{fig:11}. Each function is
normalized to unity as in Fig.~\ref{fig:5} for the unconfined packing.
The function $\rho^- (k^-)$ decreases with $k^-$ from a peak at $k^- =
0$, showing that nearly half of the particles have no tensile contact at
all. The function $\rho^+ (k^+)$ has a peak at $k^+ = 4$ and only a
small fraction $\simeq 0.05$ of particles have no compressive bond ($\rho^+
(k^+ = 0) \simeq 0.05$).

\begin{figure}
\centering
\resizebox{0.95\columnwidth}{!}{%
  \includegraphics{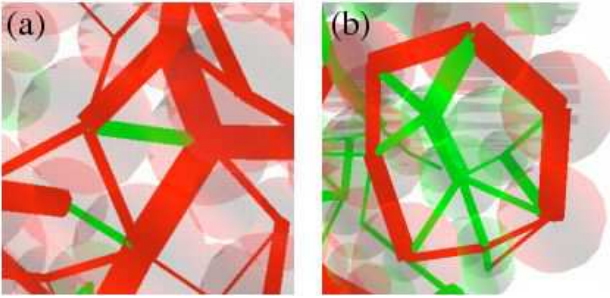}
}
\caption{(Color online) Two examples of local patterns of tensile 
(green) and compressive (red) forces surrounding particles with 
negative (green) and positive (red) pressures.}
\label{fig:12}
\end{figure}

Force patterns with one or more tensile bonds correspond to various
local configurations where equilibrium of the particles cannot be
ensured only by compressive contacts. Typically, a tensile bond between
two particles is induced by transverse particles that are forced into
the space between the two particles. One example is shown in
Fig.~\ref{fig:12}(a). For this reason, when a packing is subjected to a
stress deviator, most tensile bonds occur along the direction of
extension \cite{Richefeu06}. The upshot of tensile bonds when they are
located in a purely compressive environment is thus to reinforce the
shear strength. We also observe many particles with $2$, $3$ or $4$
tensile bonds. One example is shown in Fig.~\ref{fig:12}(b) where the
tensile actions of several particles creates a ``cage'' of compressive bonds
between their neighboring particles. This kind of patterns may be locally
self-equilibrated and thus form aggregates that could move as a rigid
body when the packing deforms.

\begin{figure}
\resizebox{0.95\columnwidth}{!}{%
  \includegraphics{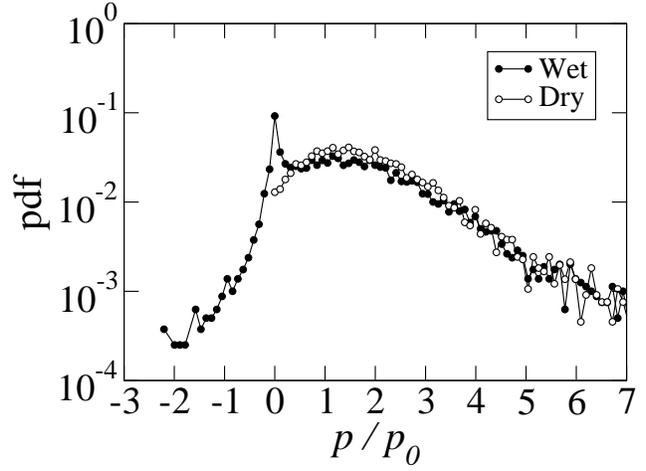}
}
\caption{Probability density function of particle pressures
normalized by a reference pressure $p_0$ (see text) in the confined packing.}
\label{fig:13}
\end{figure}

\begin{figure}
\centering
\resizebox{0.8\columnwidth}{!}{%
  \includegraphics{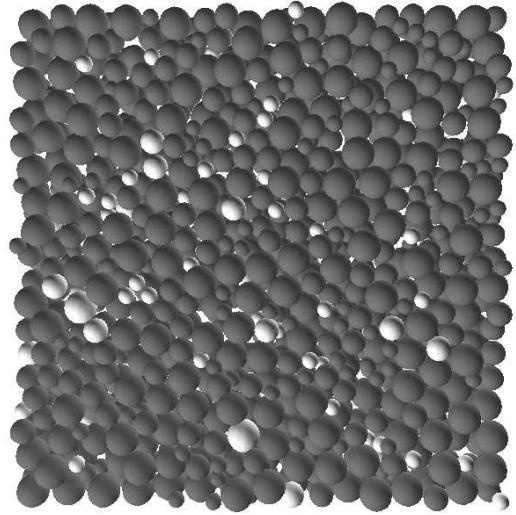}
}
\caption{A representation of a thin layer inside the confined 
packing with negative (white) and positive 
(black) particle pressures.}
\label{fig:14}
\end{figure}

The pdf of particle pressures is shown for dry and wet samples in
Fig.~\ref{fig:13}. Large positive pressures decay exponentially in both
cases. In the dry case, a local maximum is observed at $p \simeq
1.5\,p_0$ as a signature of the confining stress. In this range of
pressures, we observe a plateau for the wet sample. In the range
of negative pressures, the distribution is no more exponential. This
means that the organization of tensile forces does not fulfill the
conditions that underly exponential distribution of strong forces
in granular media \cite{Coppersmith96}. In particular, the network of tensile
forces is no more percolating throughout the packing as in the
unconfined case. A map of positive and negative particle pressures in a
thin layer inside the packing is shown in Fig.~\ref{fig:14}. The
positive pressures are dominant and negative pressure particles are
mostly isolated or appear in the form of small clusters. Although the
negative pressures do not define a bulk phase any more, the peak
centered on zero pressure in Fig.~\ref{fig:13} can still be considered as a
reminiscence of the interface between the two phases.
  
\begin{figure}
\centering
\resizebox{0.95\columnwidth}{!}{%
  \includegraphics{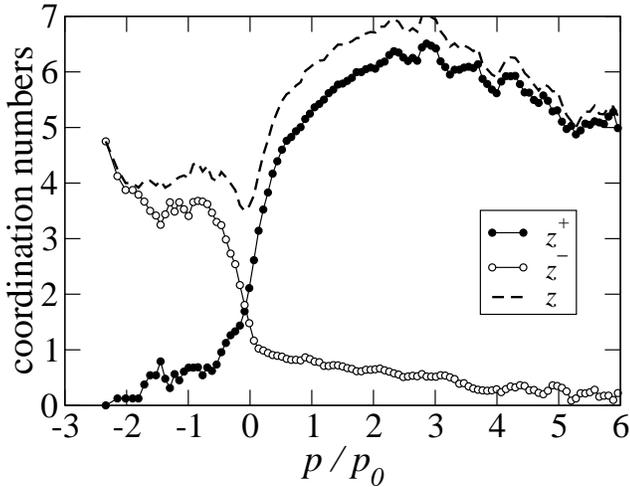}
}
\caption{Average numbers of tensile ($z^-$) and compressive ($z^+$) 
bonds per particle as well as the partial coordination number 
($z = z^- + z^+$) as a function of the particle pressure in the 
confined packing.
}
\label{fig:15}
\end{figure}

This last point appears more clearly in the plots of partial
coordination numbers $z^-$ and $z^+$ as a function of the particle
pressure $p$ in Fig.~\ref{fig:15}. We again discern three zones as
in Fig.~\ref{fig:9} for the unconfined packing. The peak $z^+ \simeq 6$
appearing around $p \simeq 2.5\,p_0$ is the effect of the
confining pressure. At the same time, the level of $z^-$ in the range of
negative pressures is reduced to $\simeq 3.5$ (from $\simeq 5$ in the
unconfined packing). The intersection occurs at $p = 0$ with $z^- = z^+
\simeq 2$.

We see that many features of stress transmission in the unconfined
packing persist when a confining stress is applied. In particular, in
both cases, a large class of particles of weak pressure (close to zero
of either sign) is present. This class was interpreted in the unconfined
case as belonging to the interface between two percolating phases.
Obviously, this interface is ill-defined for $p_m = 100$~Pa where the
negative phase appears in the form of either isolated particles or very
small aggregates. Nevertheless, our results clearly indicate that
tensile bonds and negative pressures play the same role with respect to
the equilibrium properties of the particles wherever they are present.

\begin{figure}
\centering
\resizebox{0.75\columnwidth}{!}{%
  \includegraphics{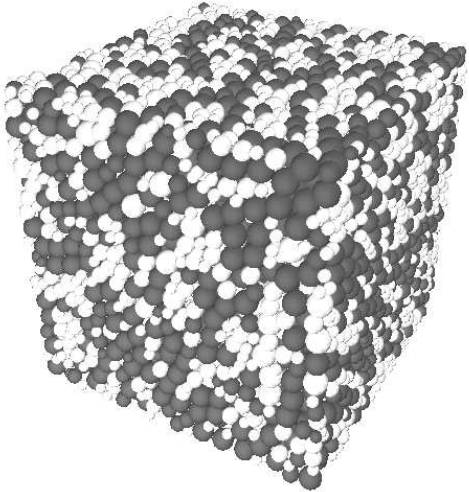}
}
\caption{A representation of the packing with particle
pressures above (white) and below (black) the mean
pressure $p_m$.}
\label{fig:16}
\end{figure}

Obviously, as stated before, the molecular dynamics method cannot be used as such
for the analysis of self-stresses in the presence of an external pressure.
This is because in molecular dynamics, the resolution of governing equations
proceeds from the knowledge of the positions of only the first neighbors 
of each particle and not from an explicit construction of the system of 
equations as in contact dynamics \cite{Moreau94,Jean99}
An approach for the analysis of self-stresses was presented 
in Radja\"i et al. \cite{Radjai96a}
using ``singular value decomposition'' in the framework of
the contact dynamics method. However, a rough estimation of the self-stresses
can be obtained by simply subtracting the mean pressure $p_m$ from the 
particle pressures. Fig.~\ref{fig:16} displays a snapshot of the particle 
pressures where the pressures below and above $p_m$ are distinguished.
The apparent clustering of particle pressures is quite comparable to
that observed in Fig~\ref{fig:8}. 
This indicates that it is very likely that if the self-stresses were isolated from 
those induced by the external pressure, they would display the same 
nearly symmetric ``bipolar'' structure as that observed at zero confining 
pressure.

\section{Conclusions}

We analyzed the statistical properties of the network of
self-equilibrated forces in a wet granular material by means of 3D
molecular dynamics simulations. Various descriptors of the
microstructure and bond force network were shown to carry the signature
of an ingenious  organization of particle pressures in two distinct
clusters of respectively positive and negative pressures, each
percolating  throughout the packing. This  partition is not meant as a
formal distinction between negative and positive pressures but rather
related to the way the two populations share the space and connect to
the bond network. This ``phase separation'' is characterized by two
interesting properties. First, the  highest pressures occur  in the
heart of each phase, whereas the lowest pressure levels constitute the
interface between the two phases. Secondly, this interface bears the
largest coordination numbers via tensile and compressive bonds. In the
presence of confining stresses, the same phenomenology  can be expected
for self-stresses although these can not be directly accessed from the
force data.        
   
It is important to remark that the homogeneity of self-stresses in our
simulations results from the homogeneous distribution of capillary
bonds. Obviously, the self-stresses can be more or less localized
in different portions of the material or involve gradients if the liquid
bonds are distributed in a nonuniform manner in the bulk.
In the same
way, although the boundary conditions  are isotropic, the self-stresses
may be locked in an anisotropic state as a result of friction and
geometrical hinderance effects. In particular, if the capillary bonds
are placed only in the gaps between particle pairs with  privileged
orientation, the self-stresses might organize into an anisotropic
scheme. Such anisotropic distributions of liquid bonds may also appear as a
result of handling the material.
The choice of a homogeneous
distribution of liquid bonds in our simulations was motivated
by the requirement of representative statistics for the analysis of
the system. However, it would be interesting to study the influence of
the liquid distribution on the patterns of self-stresses. 

The partition of self-stresses implies that the overall equilibrium of
the packing is ensured by mesoscopic structures involving length
scales larger than the particle size. These length scales are likely to
control the size of aggregates during flow or other packing properties
of cohesive granular materials. On the other hand, the effect of
self-stresses on the tensile strength or Coulomb cohesion of wet
granular materials is of interest to wet processing of grains in
chemical engineering and merits to be studied along these lines. In the
same way, the influence of solid fraction is an important aspect 
with evident application to compaction and consolidation of cohesive
packs.           

\begin{acknowledgement} 

This work was accomplished within the ``granular solids'' group of the LMGC. 
We thank F. Souli\'e for his help with the validation of the capillary law used in 
our simulations. The data were treated by means of the 
3D software \texttt{mgpost} (www.lmgc.univ-montp2.fr/$\sim$richefeu). 

\end{acknowledgement}
  
%
%

\end{document}